\documentclass[manuscript,preprint,onecolumn,aps,pre,eqsecnum,longbibliography,showpacs,showkeys,a4paper]{revtex4-1}
\usepackage[latin9]{inputenc}
\usepackage{color}
\definecolor{note_fontcolor}{rgb}{0.80078125, 0.80078125, 0.80078125}
\usepackage[english]{babel}
\usepackage{amsmath}
\usepackage{amssymb}
\usepackage{graphicx}
\usepackage{esint}
\usepackage[unicode=true,
 bookmarks=true,bookmarksnumbered=true,bookmarksopen=true,bookmarksopenlevel=1,
 breaklinks=true,pdfborder={0 0 0},backref=false,colorlinks=true]
 {hyperref}
\hypersetup{
 pdfauthor={AINS}}

\makeatletter

\newenvironment{lyxgreyedout}
  {\textcolor{note_fontcolor}\bgroup\ignorespaces}
  {\ignorespacesafterend\egroup}

\@ifundefined{textcolor}{}
{%
 \definecolor{BLACK}{gray}{0}
 \definecolor{WHITE}{gray}{1}
 \definecolor{RED}{rgb}{1,0,0}
 \definecolor{GREEN}{rgb}{0,1,0}
 \definecolor{BLUE}{rgb}{0,0,1}
 \definecolor{CYAN}{cmyk}{1,0,0,0}
 \definecolor{MAGENTA}{cmyk}{0,1,0,0}
 \definecolor{YELLOW}{cmyk}{0,0,1,0}
 }
\numberwithin{equation}{section}
\numberwithin{figure}{section}
\numberwithin{table}{section}

\usepackage{graphicx}

\DeclareGraphicsExtensions{.png,.pdf,.eps}
\graphicspath{{}}

\makeatother

\begin{document}

\title{``Systemic Nonlocality'' from Changing Constraints on Sub-Quantum
Kinematics\vspace*{\bigskipamount}
}

\author{Gerhard \surname{Grössing}\textsuperscript{1}}

\email[E-mail: ]{ains@chello.at}

\homepage[Visit: ]{http://www.nonlinearstudies.at/}

\author{Siegfried \surname{Fussy}\textsuperscript{1}}

\email[E-mail: ]{ains@chello.at}

\homepage[Visit: ]{http://www.nonlinearstudies.at/}

\author{Johannes \surname{Mesa Pascasio}\textsuperscript{1,2}}

\email[E-mail: ]{ains@chello.at}

\homepage[Visit: ]{http://www.nonlinearstudies.at/}

\author{Herbert \surname{Schwabl}\textsuperscript{1}}

\email[E-mail: ]{ains@chello.at}

\homepage[Visit: ]{http://www.nonlinearstudies.at/}

\affiliation{\textsuperscript{1}Austrian Institute for Nonlinear Studies, Akademiehof\\
 Friedrichstr.~10, 1010 Vienna, Austria}

\affiliation{\textsuperscript{2}Institute for Atomic and Subatomic Physics, Vienna
University of Technology\\
Operng.~9, 1040 Vienna, Austria\vspace*{2cm}
}

\date{\today}
\begin{abstract}
In a new approach to explain double-slit interference ``from the
single particle perspective'' via \emph{``systemic nonlocality}'',
we answer the question of how a particle going through one slit can
``know'' about the state of the other slit. We show that this comes
about by changed constraints on assumed classical sub-quantum currents,
which we have recently employed \cite{Groessing.2012doubleslit} to
derive probability distributions and Bohm-type trajectories in standard
double-slit interference on the basis of a modern, 21\textsuperscript{st}
century classical physics. Despite claims in the literature that this
scenario is to be described by a \emph{dynamical nonlocality} that
could best be understood in the framework of the Heisenberg picture
\cite{Tollaksen.2010quantum}, we show that an explanation can be
cast within the framework of the intuitively appealing Schrödinger
picture as well. We refer neither to potentials nor to a ``quantum
force'' or some other dynamics, but show that a \emph{``systemic
nonlocality''} may be obtained as a phenomenon that emerges from
an assumed sub-quantum kinematics, which is manipulated only by changing
its constraints as determined by the changes of the apparatus. Consequences
are discussed with respect to the prohibition of superluminal signaling
by standard relativity theory. %
\begin{lyxgreyedout}
\global\long\def\VEC#1{\mathbf{#1}}

\global\long\def\d{\,\mathrm{d}}

\global\long\def\e{{\rm e}}

\global\long\def\i{{\rm i}}

\global\long\def\meant#1{\left<#1\right>}

\global\long\def\meanx#1{\overline{#1}}

\global\long\def\p{\partial}
\end{lyxgreyedout}

\end{abstract}
\maketitle

\section{Introduction}

In two seminal papers \cite{Aharonov.1969modular,Aharonov.1970deterministic},
Aharonov~\emph{et\,al.\ }more than 40 years ago introduced an approach
to explain double-slit interference ``from the single particle perspective''
via \emph{dynamical nonlocality}, thereby answering the question of
how a particle going through one slit can ``know'' about the state
of the other slit (i.e., being open or closed, for example). Dynamical
nonlocality may lead to (causality preserving) changes in probability
distributions and is thus distinguished from the \emph{kinematic nonlocality}
implicit in many quantum correlations, which, however, does not cause
any changes in probability distributions.

In 2009, Tollaksen~\emph{et\,al.~}\cite{Tollaksen.2010quantum}
proposed some gedanken experiments to test the validity of this approach,
and in a recent paper by Spence and Parks first experimental evidence
for dynamical nonlocality was presented with the aid of weak measurements
\cite{Spence.2012experimental}. Essentially, it turns out that there
is a nonlocal effect of an open or a closed first slit on a particle
going through the second slit, thereby shifting the so-called modular
momentum of the particle whereas the expectation values of (moments
of) its momentum are left unchanged.

In the papers introducing dynamical nonlocality, the basis for a mechanism
to explain how a particle at the right slit can ``know'' about what
goes on at the left slit (or \emph{vice versa}) is given by the nonlocal
Heisenberg equations of motion for modular variables like the above-mentioned
modular momentum (to be discussed below). Tollaksen~\emph{et\,al.~}\cite{Tollaksen.2010quantum}
claim that one thus arrived at a fundamental difference between classical
and quantum mechanics that was ``often missed when the Schrödinger
picture is taught and classical intuitions are applied to interference.''
In contrast, we want to show here not only that a corresponding phenomenon
of\emph{ }a \emph{``systemic nonlocality''} can be accommodated
within the Schrödinger picture, too, but also that a completely ``classical''
approach is feasible if one understands ``classical'' in the sense
of present-day, ``21\textsuperscript{st} century classical physics'',
i.e., including recent developments such as diffusion wave fields,
superstatistics, or ballistic diffusion, for example. It is clear
that both the Heisenberg and the Schrödinger pictures, respectively,
have their advantages and disadvantages, for example w.r.t.~applications
to more complex scenarios, but they also complement each other, thereby
highlighting different aspects with different useful insights. In
this sense, our intention is not to criticize the use of the Heisenberg
picture for the problem in question, but to show the usefulness of
the Schrödinger picture as well, thus throwing light on ``systemic
nonlocality'' from a perspective which is more in line with common
intuitions.

In \cite{Tollaksen.2010quantum,Aharonov.1969modular,Aharonov.1970deterministic},
the Heisenberg equations of motion are shown to be nonlocal because
of the dependence on the potential at two distinct locations (i.e.,
of the slits). Modeled like the scalar Aharonov-Bohm effect, one finds
also here that there are no forces acting on the particle. For example,
when a particle reaches a slit and the other slit is suddenly closed,
the authors claim that the only effect is a nonlocal change of the
particle's modular momentum due to the associated change of the potential
at the location of the other (now closed) slit.

In this paper, we substitute the description via potentials in double
slit interference by considering only constraints, or boundary conditions,
on an assumed sub-quantum kinematics. Thus, putting a hypothesized
sub-quantum medium into a systemic context, we shall attempt to model
``systemic nonlocality'' not via changes of potential differences,
but by changes in sub-quantum kinematics due to their changed constraints.
This amounts to a conceptual switch from a formal consideration of
the macroscopic potentials of the experimental setup to a more ``material''
consideration of microscopic, i.e., sub-quantum currents whose modifications
upon changing boundary conditions will be shown to co-determine the
``nonlocal effects''. (The reason for using the terms such as ``systemic
nonlocality'' or ``nonlocal effects'' only within quotation marks
is given by the circumstance, to be discussed in Section 4, that these
descriptions are supposed to only hold for the time resolutions of
present-day experiments, but to eventually break down at very small
time scales.) Just as in the above-quoted papers on dynamical nonlocality,
we shall obtain the usual quantum mechanical results by introducing
no potential-related forces whatsoever. Although our approach has
many features in common with a Bohmian one, this is where an essential
difference must be pointed out. Whereas in Bohmian theory, a nonlocal
``quantum force'' is made responsible for the genuine quantum effects,
in our model no such force exists, since the equivalent of the Bohmian
quantum potential is given by contributions from purely kinetic energy
terms. As we employ no ``quantum force'', therefore, we consider
\emph{``systemic nonlocality'' as a phenomenon that emerges from
a sub-quantum kinematics}, which is manipulated only by changing its
constraints as determined by the changes of the apparatus.

In fact, with our approach we have in a series of papers obtained
essential elements of quantum theory \cite{Groessing.2012doubleslit,Groessing.2008vacuum,Groessing.2009origin,Groessing.2010emergence,Groessing.2011dice,Groessing.2011explan,Grossing.2012quantum}.
They derive from the assumption that a particle of energy $E=\hbar\omega$
is actually an oscillator of angular frequency $\omega$ phase-locked
with the zero-point oscillations of the surrounding environment, the
latter of which containing both regular and fluctuating components
and being constrained by the boundary conditions of the experimental
setup via the buildup and maintenance of standing waves. The particle
in this approach is an off-equilibrium steady-state maintained by
the throughput of zero-point energy from its vacuum surroundings.
This is in close analogy to the bouncing/walking droplets in the experiments
of Couder's group \cite{Couder.2005,Couder.2006single-particle,Couder.2010walking,Couder.2012probabilities},
which in many respects can serve as a classical prototype guiding
our intuition. We have recently applied our model to the case of interference
at a double slit \cite{Groessing.2012doubleslit}, thereby obtaining
the exact quantum mechanical probability distributions on a screen
behind the double slit, the average particle trajectories (which because
of the averaging are shown to be identical to the Bohmian ones), and
the involved probability density currents. One aim of the present
paper is to extend the applicability of our model to a scenario that
is only apparently dynamical, i.e., to describe ``systemic nonlocality''
as introduced above.

This paper is structured as follows. In Section~\ref{sec:A-short-Introduction},
we give a short introduction to our sub-quantum model, with explanations
of Gaussian dispersion and interference at a double slit, among others.
Section~\ref{sec:Dynamical-Nonlocality-in} presents a review of
dynamical nonlocality, its explanations in the recent literature via
the Heisenberg picture, and our own explanation via the Schrödinger
picture. Finally, in Section~\ref{sec:Emergence-of-Dynamical} we
show how a corresponding ``systemic nonlocality'' can be understood
as emerging from a classical sub-quantum kinematics.

\section{A Short Introduction to our Sub-Quantum Approach to Quantum Mechanics\label{sec:A-short-Introduction}}

In older approaches to quantum theory by de\,Broglie or Bohm, for
example, one basically focused on a role of the waves, although sometimes
possibly ``empty'', as essentially ``guiding'' the particle through
the experimental apparatus. However, the Couder experiments explicitly
point at a more complex scenario by an observed partial decoupling
of wave and particle propagation. While the particle (or bouncer/walker,
respectively) still is guided through interfering and fluctuating
waves, the latter constitute a landscape that is not only present
in the vicinity of the particle, but throughout the \emph{whole apparatus}.
(In a rare illustration of a similar circumstance in quantum physics,
Bohm and Hiley write: \textquotedblleft{}\ldots{}it is only through
the existence of\ldots{} pools of information which are not expressible
solely in terms of relationships of actual particles that the notion
of an objective whole can be given meaning.\textquotedblright{} \cite{Bohm.1993undivided})

We transfer the corresponding insight from the Couder experiments
into our modeling of quantum systems and assume that the waves are
a space-filling phenomenon involving the whole experimental setup.
Thus, one can imagine a partial decoupling of the physics of waves
and particles in that the latter still may be ``guided'' through
said landscape, but the former may influence other regions of the
landscape by providing specific phase information independently of
the propagation of the particle. This is why a remote change in the
experimental setup, when mediated to the particle via de- and/or re-construction
of standing waves, can potentially amount to a nonlocal effect on
a particle via the thus modified guiding landscape.

In \cite{Groessing.2010emergence} we presented a model for the classical
explanation of the quantum mechanical dispersion of a free Gaussian
wave packet. In accordance with the classical model, we shall now
relate it more directly to a ``double solution'' analogy gleaned
from Couder and Fort \cite{Couder.2012probabilities}. For, as is
shown, e.g., in \cite{Holland.1993}, one can construct various forms
of classical analogies to the quantum mechanical Gaussian dispersion.
Originally, the expression of a ``double solution'' refers to an
early idea of de\,Broglie \cite{DeBroglie.1960book} to model quantum
behavior by a two-fold process, i.e., by the movement of a hypothetical
point-like singularity solution of the Schrödinger equation, and by
the evolution of the usual wavefunction that would provide the empirically
confirmed statistical predictions. Recently, Couder and Fort \cite{Couder.2012probabilities}
used this ansatz to describe the behaviors of their bouncer- (or walker-)
droplets: on an individual level, one observes particles surrounded
by circular waves they emit through the phase-coupling with an oscillating
bath, which provides, on a statistical level, the emergent outcome
in close analogy to quantum mechanical behavior (like, e.g., diffraction
or double-slit interference).

In the context of the double solution idea, which is related to correlations
on a statistical level between individual uncorrelated particle positions
$x$ and momenta $p$, respectively, we consider the free Liouville
equation. It provides a phase-space distribution $f\left(x,p,t\right)$
that shows the emergence of correlations between $x$ and $p$ from
an initially uncorrelated product function of non-spreading (classical)
Gaussian position distributions as well as momentum distributions.
The motivation for their introduction comes exactly from what one
observes in the Couder experiments. In an idealized scenario, we assume
that at each point $x$ an unbiased emission of momentum fluctuations
$\pi_{0}$ in all possible directions takes place, thus mimicking
(in a two-dimensional scenario) the circular waves emitted from the
``particle as bouncer''. If we compare the typical frequency of
the bouncers in the Couder experiments (i.e., roughly $10^{2}\mathrm{\, Hz}$)
with that of an electron, for example (i.e., roughly $10^{20}\,\mathrm{Hz}$),
we see that a continuum ansatz is practically plausible, particularly
if we are interested in statistical averages over a long series of
experimental runs. 

Thus, one can construct said phase-space distribution, with $\sigma_{0}$
being the initial $x$--space standard deviation, i.e., $\sigma_{0}=\sigma(t=0)$,
and $\pi_{0}:=mu_{0}$ the momentum standard deviation, such that
\begin{equation}
f\left(x,p,t\right)=\frac{1}{2\pi\sigma_{0}mu_{0}}\exp\left\{ -\frac{\left(x-pt/m\right)^{2}}{2\sigma_{0}^{2}}\right\} \exp\left\{ -\frac{p^{2}}{2m^{2}u_{0}^{2}}\right\} .\label{eq:traj.2.6}
\end{equation}

\noindent Now, the above-mentioned correlations between $x$ and $p$
emerge when one considers the probability density in $x$--space.
It turns out that
\begin{equation}
P\left(x,t\right)=\int f\d p=\frac{1}{\sqrt{2\pi}\sigma}\exp\left\{ -\frac{x^{2}}{2\sigma^{2}}\right\} ,\label{eq:traj.2.7}
\end{equation}
with the standard deviation at time $t$ given by
\begin{equation}
\sigma^{2}=\sigma_{0}^{2}+u_{0}^{2}\, t^{2}.\label{eq:traj.2.8}
\end{equation}

In other words, the distribution (\ref{eq:traj.2.7}) with (\ref{eq:traj.2.8})
describing a spreading Gaussian is obtained from a continuous set
of classical, i.e., non-spreading, Gaussian position distributions
of particles whose associated momentum fluctuations also have non-spreading
Gaussian distributions. One thus obtains the exact quantum mechanical
dispersion formula for a Gaussian, as we have obtained also previously
from our classical ansatz. For confirmation with respect to our diffusion
model \cite{Groessing.2010emergence,Groessing.2011dice}, we note
that with the usual definition of the ``osmotic'' velocity field
$u=-D\frac{\nabla P}{P}$, one obtains with (\ref{eq:traj.2.7}),
and with bars denoting averages,
\begin{equation}
\meanx{u^{2}}=D^{2}\meanx{\left(\frac{\nabla P}{P}\right)^{2}}=\frac{D^{2}}{\sigma^{2}},\qquad\textrm{and thus also}\qquad u_{0}=\frac{D}{\sigma_{0}},\label{eq:traj.2.9}
\end{equation}

\noindent so that one can rewrite Eq.~(\ref{eq:traj.2.8}) in the
more familiar form
\begin{equation}
\sigma^{2}=\sigma_{0}^{2}\left(1+\frac{D^{2}t^{2}}{\sigma_{0}^{4}}\right).\label{eq:traj.2.10}
\end{equation}

\noindent Note also that by using the Einstein relation $D=\hbar/(2m)=h/(4\pi m)$
the norm in (\ref{eq:traj.2.6}) thus becomes the invariant expression
(reflecting the ``exact uncertainty relation'' \cite{Hall.2002schrodinger})
\begin{equation}
\frac{1}{2\pi\sigma_{0}mu_{0}}=\frac{1}{2\pi mD}=\frac{2}{h}.\label{eq:traj.2.11}
\end{equation}

Following from (\ref{eq:traj.2.10}), in references \cite{Groessing.2012doubleslit,Groessing.2010emergence}
we obtained for smoothed-out trajectories (i.e., averaged over a very
large number of Brownian motions) a sum over an Ehrenfest-type and
a fluctuations term, respectively, for the motion in the $x$--direction
\begin{equation}
x_{{\rm tot}}(t)=vt+x(t)=vt+x(0)\frac{\sigma}{\sigma_{0}}=vt+x(0)\sqrt{1+\frac{D^{2}t^{2}}{\sigma_{0}^{4}}}.\label{eq:2.22}
\end{equation}
Thus one obtains the \emph{average velocity field} of a Gaussian wave
packet as 
\begin{equation}
v_{{\rm tot}}(t)=v(t)+\frac{\d x(t)}{{\rm \d t}}=v(t)+\left[x_{{\rm tot}}(t)-vt\right]\frac{u_{0}^{2}t}{\sigma^{2}}.\label{eq:2.23}
\end{equation}

Note that Eqs.~(\ref{eq:2.22}) and (\ref{eq:2.23}) are derived
solely from statistical physics. Still, they are in full accordance
with quantum theory, and in particular with Bohmian trajectories \cite{Holland.1993}.
Note also that one can rewrite Eq.~(\ref{eq:traj.2.10}) such that
it appears like a linear-in-time formula for Brownian motion, 
\begin{equation}
\meanx{x^{2}}=\meanx{x^{2}(0)}+D_{t}\, t,\label{eq:2.24}
\end{equation}
 where a time dependent diffusivity 
\begin{equation}
D_{\mathrm{t}}=u_{0}^{2}\, t=\frac{\hbar^{2}}{4m^{2}\sigma_{0}^{2}}\, t\label{eq:2.25}
\end{equation}
characterizes Eq.~(\ref{eq:2.24}) as \textit{ballistic diffusion}.
This makes it possible to simulate the dispersion of a Gaussian wave
packet on a computer by simply employing coupled map lattices for
classical diffusion, with the diffusivity given by Eq.~(\ref{eq:2.25}).
(For detailed discussions, see refs.~\cite{Groessing.2010emergence}
and \cite{Groessing.2011dice}.)

Moreover, one can easily extend this scheme to more than one slit,
like, for example, to explain interference effects at the double slit\cite{Groessing.2012doubleslit,Grossing.2012quantum}.
For this, we chose similar initial situations as in \cite{Holland.1993},
i.e., electrons (represented by plane waves in the forward $y$--direction)
from a source passing through soft-edged slits $1$ and $2$ in a
barrier (located along the $x$--axis) and recorded at a screen. In
our model, we therefore note two Gaussians representing the totality
of the effectively ``heated-up'' so-called \emph{path excitation
field} (to be detailed below), one for slit $1$ and one for slit
$2$, whose centers have the distances $+X$ and $-X$ from the plane
spanned by the source and the center of the barrier along the $y$--axis,
respectively. 

With the total amplitude $R$ of two coherent waves with (suitably
normalized) amplitudes $R_{i}=\sqrt{P_{i}}$, and the local phases
$\varphi_{i}$, $i=1$ or $2$, one has as usual that
\begin{equation}
R=R_{1}\cos\left(\omega t+\varphi_{1}\right)+R_{2}\cos\left(\omega t+\varphi_{2}\right).\label{eq:11a}
\end{equation}
Introducing an arbitrarily chosen unit vector $\mathbf{\hat{n}}$,
one may also define $\cos\left(\omega t+\varphi_{i}\left(\mathbf{x}\right)\right)$=
$\mathbf{\hat{n}\cdot\mathbf{\hat{k_{i}}\left(\mathbf{x}\mathrm{,t}\right)}},$
such that along with the system's evolution, the emergent outcome
of the time evolution of (\ref{eq:11a}) can be written as 
\begin{equation}
R\left(\mathbf{x}\mathrm{,t}\right)=\hat{\mathbf{n}}\cdot\left(R_{1}\mathbf{\hat{k}}_{1}\left(\mathbf{x}\mathrm{,t}\right)+R_{2}\mathbf{\hat{k}}_{2}\left(\mathbf{x}\mathrm{,t}\right)\right),\label{eq:11b}
\end{equation}
which we shall use later on. According to classical textbook wisdom,
the \emph{averaged total intensity }becomes
\begin{equation}
P_{{\rm tot}}:=R^{2}=R_{1}^{2}+R_{2}^{2}+2R_{1}R_{2}\cos\varphi_{12}=P_{1}+P_{2}+2\sqrt{P_{1}P_{2}}\cos\varphi_{12},\label{eq:3.0a-1-1}
\end{equation}
where $\varphi_{12}$ is the relative phase $\varphi_{12}=\varphi_{1}-\varphi_{2}=\left(\VEC k_{1}-\VEC k_{2}\right)\cdot\VEC r$.
Note that $\varphi_{12}$ enters Eq.~(\ref{eq:3.0a-1-1}) only via
the cosine function, such that, e.g., even if the total wave numbers
(and thus also the total momenta) $\VEC k_{i}$ were of vastly different
size, the cosine effectively makes Eq.~(\ref{eq:3.0a-1-1}) independent
of said sizes, but dependent only on an angle modulo $2\pi$. This
will turn out as essential for our discussion further below.

The $x$--components of the centroids' motions from the two alternative
slits $1$ and $2$, respectively, are given by the particle velocity
components 
\begin{equation}
v_{x}=\pm\frac{\hbar}{m}\, k_{x},\label{eq:3.1-1-1}
\end{equation}
 respectively, such that the relative group velocity of the Gaussians
spreading into each other is given by $\Delta v_{x}=2v_{x}$. However,
in order to calculate the phase difference $\varphi_{12}$ descriptive
of the interference term of the intensity distribution (\ref{eq:3.0a-1-1}),
one must take into account the total momenta involved, i.e., one must
also include the wave packet dispersion as described in the previous
Section. Thus, one obtains with the displacement $\pm x\left(t\right)=\mp\left(X+v_{x}t\right)$
in Eq.~(\ref{eq:2.23}) the total relative velocity of the two Gaussians
as 
\begin{equation}
\Delta v_{{\rm tot},x}=2\left[v_{x}-(X+v_{x}t)\frac{u_{0}^{2}t}{\sigma^{2}}\right].\label{eq:3.1a-1-1}
\end{equation}
Therefore, the total phase difference between the two possible paths
1 and 2 (i.e., through either slit) becomes 
\begin{equation}
\varphi_{12}=\frac{1}{\hbar}(m\Delta v_{{\rm tot},x}\, x)=2mv_{x}\frac{x}{\hbar}-(X+v_{x}t)x\frac{1}{D}\frac{u_{0}^{2}t}{\sigma^{2}}.\label{eq:dyn.2.19}
\end{equation}

In our earlier papers \cite{Groessing.2008vacuum,Groessing.2009origin,Groessing.2010emergence},
we have shown that, apart from the ordinary particle current $\VEC J(\VEC x,t)=P(\VEC x,t)\VEC v$,
we are now dealing with two additional, yet opposing, currents $\VEC J_{u}=P(\VEC x,t)\VEC u$,
which are on average orthogonal to $\VEC J$ \cite{Groessing.2008vacuum,Groessing.2009origin,Groessing.2010emergence,Groessing.2011explan},
and which are the emergent outcome from the presence of numerous corresponding
velocities 
\begin{equation}
\VEC u_{\pm}=\mp\frac{\hbar}{2m}\frac{\nabla P}{P}.\label{eq:2.7}
\end{equation}

We denote with $\VEC u_{+}$ and $\VEC u_{-}$, respectively, the
two opposing tendencies of the diffusion process. Moreover, when we
take the averages, we obtain a smoothed-out \textit{average velocity
field} 
\begin{equation}
\meanx{\VEC u(\VEC x,t)}=\int P\VEC u(\VEC x,t)\d^{n}x,\label{eq:2.10}
\end{equation}
which is all that we need for our further considerations. Similarly,
based on the fact that we have an initial Gaussian distribution of
velocity vectors $\VEC v(\VEC x,t)$, we define an average velocity
field $\meanx{\VEC v}$ of the wave propagation as 
\begin{equation}
\meanx{\VEC v(\VEC x,t)}=\int P\VEC v(\VEC x,t)\d^{n}x,\label{eq:2.11}
\end{equation}
and for the free particle make use of an average orthogonality between
the two velocity fields, $\VEC u$ and $\VEC v$, \cite{Groessing.2008vacuum,Groessing.2009origin,Groessing.2010emergence,Groessing.2011dice},
\begin{equation}
\meanx{\VEC v\cdot\VEC u}=\int P\VEC v\cdot\VEC u\d^{n}x=0.\label{eq:2.12}
\end{equation}

In effect, then, the combined presence of both velocity fields $\VEC u$
and $\VEC v$ can be denoted as a \textit{path excitation field}:
via diffusion, the bouncer in its interaction with already existing
wave-like excitations of the environment creates an ``agitated'',
or ``heated-up'', thermal landscape, which can also be pictured
by interacting wave configurations all along between source and detector
of an experimental setup. Recall that our prototype of a walking bouncer,
i.e., from the experiments of Couder's group, is always driven by
its interactions with a superposition of waves emitted at the points
it visited in the past. Couder~\emph{et\,al}.\ denote this superposition
of in-phase waves the ``path memory'' of the bouncer \cite{Fort.2010path-memory}.
This implies, however, that the bouncers at the points visited in
the present necessarily create new wave configurations which will
form the basis of a path memory in the future. In other words, the
wave configurations of the past determine the bouncer's path in the
present, whereas its bounces in the present co-determine the wave
configurations at any of the possible locations it will visit in the
future. Therefore, we call the latter configurations the \textit{path
excitation field}, which may also be described as heated-up thermal
field. Ideally, as in the coupling of an oscillator with classical
diffusion, non-relativistic diffusion wave fields arise with instantaneous
field propagation \cite{Groessing.2009origin,Mandelis.2001structure},
one has elements of the whole setup which may be nonlocally oscillating
(``breathing'') in phase. This means that the Gaussian of Eq.~(\ref{eq:traj.2.7})
can be said to represent an idealized nonlocal path excitation field
in that it is a physically existing and effective entity responsible
for where the bouncing particle can possibly go. (We shall discuss
a less idealized approach in the last Section of this paper.) 

Let us now consider a single, classical particle (bouncer) following
the propagation of a set of waves of equal amplitude $R_{i}$, each
representing one of $i$ possible alternatives according to our principle
of path excitation, and focus on the specific role of the velocity
fields. To describe the required details, each path $i$ be occupied
by a Gaussian wave packet with a forward momentum $\VEC p_{i}=\hbar\VEC k_{i}=m\VEC v_{i}$.
Moreover, due to the stochastic process of path excitation, the latter
has to be represented also by a large number $N$ of consecutive Brownian
shifts, $\VEC p_{u,\alpha}=m\VEC u_{\alpha}$. Defining (with indices
$i=1$ or $2$ referring to the two slits, and with $+$ and $-$
referring to the right and the left from the average direction of
$\VEC v_{i}$, respectively) 
\begin{equation}
\meanx{\VEC v}_{{\rm tot},i}:=\meanx{\VEC v}_{i}+\meanx{\VEC u}_{i+}+\meanx{\VEC u}_{i-,}\label{eq:3.8}
\end{equation}
and with two Gaussian distributions $P_{1}=R_{1}^{2}$ and $P_{2}=R_{2}^{2}$,
one has with (\ref{eq:11b}) 
\begin{equation}
R_{{\rm tot}}^{2}=\left(R_{1}\meanx{\VEC{\hat{v}}}_{{\rm tot},1}+R_{2}\meanx{\VEC{\hat{v}}}_{{\rm tot},2}\right)^{2}\;,\label{eq:3.9}
\end{equation}
which after a few calculational steps provides (similarly to \cite{Groessing.2012doubleslit},
but now with slightly different labellings which apply more generally)
the total average current 
\begin{equation}
\meanx{\VEC J}_{{\rm tot}}=P_{1}\meanx{\VEC v}_{1}+P_{2}\meanx{\VEC v}_{2}+\sqrt{P_{1}P_{2}}\left(\meanx{\VEC v}_{1}+\meanx{\VEC v}_{2}\right)\cos\varphi_{12}+\sqrt{P_{1}P_{2}}\left(\meanx{\VEC u}_{1}-\meanx{\VEC u}_{2}\right)\sin\varphi_{12}.\label{eq:3.12}
\end{equation}
A more detailed account of Eq.~(\ref{eq:3.12}) and its extension
to \emph{n} slits is in preparation {[}Fussy~\emph{et\,al}.\ (2013){]}.
Note that Eq.~(\ref{eq:3.12}), upon the identification of $\meanx{\VEC u}_{i}=-\frac{\hbar}{m}\frac{\nabla R_{i}}{R_{i}}$
from Eq.~(\ref{eq:2.7}) and with $P_{i}=R_{i}^{2}$, turns out to
be in perfect agreement with a comparable ``Bohmian'' derivation~\cite{Holland.1993,Sanz.2008trajectory}.
In fact, with $\meanx{\VEC v}_{i}=$$\frac{\nabla S_{i}}{m}$, one
can rewrite (\ref{eq:3.12}) as
\begin{equation}
\meanx{\VEC J}_{{\rm tot}}=R_{1}^{2}\frac{\nabla S_{1}}{m}+R_{2}^{2}\frac{\nabla S_{2}}{m}+R_{1}R_{2}\left(\frac{\nabla S_{1}}{m}+\frac{\nabla S_{2}}{m}\right)\cos\varphi_{12}+\frac{\hbar}{m}\left(R_{1}\nabla R_{2}-R_{2}\nabla R_{1}\right)\sin\varphi_{12}.\label{eq:dyn.3.12.1}
\end{equation}

The formula for the averaged particle trajectories, then, simply results
from 
\begin{equation}
\meanx{\VEC v}_{{\rm tot}}=\frac{\meanx{\VEC J}_{{\rm tot}}}{P_{\textrm{tot}}}.\label{eq:3.13}
\end{equation}

Although we have obtained the usual quantum mechanical results, we
have so far not used the quantum mechanical formalism in any way.
However, upon employment of the Madelung transformation for each path
$j$ ($j=1$ or $2$), 
\begin{equation}
\Psi_{j}=R\e^{\i S/\hbar},\label{eq:3.14}
\end{equation}
 and thus $P_{j}=R_{j}^{2}=|\Psi_{j}|^{2}=\Psi_{j}^{*}\Psi_{j}$,
with the definitions (\ref{eq:2.7}) and $v_{j}:=\nabla S_{j}/m$,
$\varphi_{12}=(S_{1}-S_{2})/\hbar$, and recalling the usual trigonometric
identities such as $\cos\varphi=\frac{1}{2}\left(\e^{\i\varphi}+\e^{-\i\varphi}\right)$,
etc., one can rewrite the total average current (\ref{eq:3.12}) immediately
as 
\begin{equation}
\begin{array}{rl}
{\displaystyle \meanx{\VEC J}_{{\rm tot}}} & =P_{{\rm tot}}\meanx{\VEC v}_{{\rm tot}}\\[3ex]
 & ={\displaystyle (\Psi_{1}+\Psi_{2})^{*}(\Psi_{1}+\Psi_{2})\frac{1}{2}\left[\frac{1}{m}\left(-\i\hbar\frac{\nabla(\Psi_{1}+\Psi_{2})}{(\Psi_{1}+\Psi_{2})}\right)+\frac{1}{m}\left(\i\hbar\frac{\nabla(\Psi_{1}+\Psi_{2})^{*}}{(\Psi_{1}+\Psi_{2})^{*}}\right)\right]}\\[3ex]
 & ={\displaystyle -\frac{\i\hbar}{2m}\left[\Psi^{*}\nabla\Psi-\Psi\nabla\Psi^{*}\right]={\displaystyle \frac{1}{m}{\rm Re}\left\{ \Psi^{*}(-\i\hbar\nabla)\Psi\right\} ,}}
\end{array}\label{eq:3.18}
\end{equation}
 where $P_{{\rm tot}}=|\Psi_{1}+\Psi_{2}|^{2}=:|\Psi|^{2}$. The last
two expressions of (\ref{eq:3.18}) are the exact well-known formulations
of the quantum mechanical probability current, here obtained without
any quantum mechanics, but just by a re-formulation of (\ref{eq:3.12}).
In fact, it is a simple exercise to insert the wave functions (\ref{eq:3.14})
into (\ref{eq:3.18}) to re-obtain (\ref{eq:3.12}).

\section{``Systemic'' versus Dynamical Nonlocality in Double Slit Interference\label{sec:Dynamical-Nonlocality-in}}

In reference \cite{Tollaksen.2010quantum}, Tollaksen~\emph{et\,al}.\ discuss
interference at a double slit ``from a single particle perspective''
and ask the following question: If a particle goes through one slit,
how does it ``know'' whether the second slit is open or closed?
We shall here first recapitulate the arguments providing these authors'
answer and later provide our own arguments and answer. Of course,
the question is about the phase information and how it affects the
particle. We know from quantum mechanics that phases cannot be observed
on a local basis and that a common overall phase has no observational
meaning. Assuming that two initially non-overlapping Gaussian wave
functions, $\Psi_{1}$ and $\Psi_{2}$, describe the probability amplitudes
for particles emerging from slits $1$ or $2$, respectively, which
are separated by a distance $D,$ the total wave function for the
particle exiting the double slit may be written as
\begin{equation}
\Psi=\e^{\i\alpha_{1}}\Psi_{1}+\e^{\i\alpha_{2}}\Psi_{2},
\end{equation}
but since a common overall phase is insignificant, one writes the
total wave function as
\begin{equation}
\Psi_{\varphi}=\Psi_{1}+\e^{\i\varphi}\Psi_{2},
\end{equation}
where $\varphi:=\alpha_{2}-\alpha_{1}$ is the physically significant
\emph{relative phase}. Tollaksen~\emph{et\,al}.\ now ask where
the relative phase appears in the form of (deterministic) observables
that describe interference. When looking at the expectation value
(in the one-dimensional case, for simplicity)
\begin{align}
\overline{x}=\left[\int\Psi_{\varphi}^{*}\Psi_{\varphi}\d x\right]^{-1}\int\Psi_{\varphi}^{*}x\Psi_{\varphi}\d x & \equiv\int\left(\Psi_{1}^{*}+\e^{-\i\varphi}\Psi_{2}^{*}\right)x\left(\Psi_{1}+\e^{\i\varphi}\Psi_{2}\right)\d x\\
 & =\int\left(\Psi_{1}^{*}x\Psi_{1}+\Psi_{2}^{*}x\Psi_{2}\right)\d x+\int\Psi_{1}^{*}x\e^{\i\varphi}\Psi_{2}\d x+\mathrm{c.c.},\nonumber 
\end{align}
one sees that it is independent of $\varphi$ because of the vanishing
of the last term due to the assumed non-overlapping of $\Psi_{1}$
and $\Psi_{2}$. Similarly, this also holds for all moments $\overline{x^{n}},$
and for all moments $\overline{p^{n}}$ as well. In particular, one
has for the expectation value of the momentum
\begin{align}
\overline{p} & =\left[\int\Psi_{\varphi}^{*}\Psi_{\varphi}\d x\right]^{-1}\mathrm{Re}\int\Psi_{\varphi}^{*}\i\hbar\frac{\partial}{\partial x}\Psi_{\varphi}\d x\label{eq:dyn.3.1-0}\\
 & \equiv\mathrm{Re}\int\left(\Psi_{1}^{*}\i\hbar\frac{\partial}{\partial x}\Psi_{1}+\Psi_{2}^{*}\i\hbar\frac{\partial}{\partial x}\Psi_{2}\right)\d x+\mathrm{Re}\int\Psi_{1}^{*}\i\hbar\frac{\partial}{\partial x}\e^{\i\varphi}\Psi_{2}\d x+\mathrm{c.c.},\nonumber 
\end{align}
where the $\varphi$--dependent term vanishes identically, because
$\frac{\partial}{\partial x}\Psi_{2}=0$ for $\Psi_{2}=0$. So, again,
where does the relative phase appear? The answer of Tollaksen\emph{~et\,al}.\ is
given by a ``shift operator'' that shifts the location of, say $\Psi_{1}$,
over the distance $D$ to its new location coinciding with that of
$\Psi_{2}$. The expectation value of the shift operator is thus given
by
\begin{equation}
\overline{\e^{-\i\frac{pD}{\hbar}}}\equiv\int\left(\Psi_{1}^{*}\e^{-\i\frac{pD}{\hbar}}\Psi_{1}+\Psi_{2}^{*}\e^{-\i\frac{pD}{\hbar}}\Psi_{2}\right)\d x+\int\Psi_{1}^{*}\e^{-\i\frac{pD}{\hbar}}\Psi_{2}\e^{\i\varphi}\d x+\int\Psi_{2}^{*}\e^{-\i\varphi}\e^{-\i\frac{pD}{\hbar}}\Psi_{1}\d x,
\end{equation}
where all but the last term vanish identically, thus providing (with
the correct normalization)
\begin{equation}
\overline{\e^{-\i\frac{pD}{\hbar}}}=\e^{-\i\varphi}/2\;\textrm{ and }\;\overline{\e^{-\i\frac{pD}{\hbar}}}+\overline{\e^{\i\frac{pD}{\hbar}}}=\cos\varphi.
\end{equation}
In order to make sense of the shift operator, the authors now shift
to the Heisenberg picture, thereby providing with a Hamiltonian $H=\frac{p^{2}}{2m}+V\left(x\right)$
the time evolution of the operator as
\begin{equation}
\frac{\d}{\d t}\:\e^{-\i\frac{pD}{\hbar}}=\frac{\i}{\hbar}\left[\e^{-\i\frac{pD}{\hbar}},\frac{p^{2}}{2m}+V\left(x\right)\right]=\frac{-\i D}{\hbar}\e^{-\i\frac{pD}{\hbar}}\left\{ \frac{V\left(x\right)-V\left(x+D\right)}{D}\right\} .\label{eq:dyn.3.1-1}
\end{equation}
With its dependence on the distance $D$ between the two slits, Eq.
(\ref{eq:dyn.3.1-1}) is claimed to be a description of \emph{dynamical
nonlocality} showing how a particle can ``know'' about the presence
of the other slit. Tollaksen\emph{~et\,al}.\ maintain that it is
possible to understand this dynamical nonlocality only by employing
the Heisenberg picture. However, we shall now show that an equivalent
understanding is possible also within the Schrödinger picture, and
even more intuitively accessible, too. For, there is one assumption
in the foregoing analysis that is not guaranteed to hold in general,
i.e., the non-overlapping of the wave functions $\Psi_{1}$ and $\Psi_{2}$.
On the contrary, we now shall assume that the two Gaussians representing
the probability amplitudes for the particle immediately after passing
one of the two slits do not have any artificial cut-off, but actually
extend across the whole slit system, even with only very small (and
practically often negligible) amplitudes in the regions further away
from the slit proper. (We shall give arguments for this assumption
further below.) In other words, we now ask: what if $\Psi_{1}$ and
$\Psi_{2}$ do overlap, even if only by a very small amount? To answer
this question, we consider the expectation value of the momentum,
or the total current $\mathbf{J}_{\mathrm{tot}}=P_{\mathrm{tot}}\frac{\bar{\mathbf{p}}}{m}$,
respectively, and we obtain from Eq.~(\ref{eq:dyn.3.1-0}) that the
terms $\mathbf{p}_{\varphi}$ involving the relative phase $\varphi$
stem from the interference terms of $\mathbf{J}_{\mathrm{tot}}$,
i.e., they provide with $\Psi_{j}=R_{j}e^{\i\varphi_{j}}$ and $\varphi_{j}=\frac{S_{j}}{\hbar}$
\begin{equation}
P_{\mathrm{tot}}\frac{\overline{\mathbf{p}_{\varphi}}}{m}=R_{1}R_{2}\left(\frac{\nabla S_{1}}{m}+\frac{\nabla S_{2}}{m}\right)\cos\varphi+\frac{\hbar}{m}\left(R_{1}\nabla R_{2}-R_{2}\nabla R_{1}\right)\sin\varphi.\label{eq:dyn.3.1-2}
\end{equation}
First of all one notes upon comparison of Eq.~(\ref{eq:dyn.3.1-2})
with Eqs.~(\ref{eq:3.12})--(\ref{eq:3.13}) the exact correspondence
of (\ref{eq:dyn.3.1-2}) with our classically obtained expression
for the interference terms of the emerging current, or of the expression
for $\mathbf{\overline{v}}_{\mathrm{tot}}$, respectively. Moreover,
although the product $R_{1}R_{2}$ is in fact negligibly small for
regions where only a long tail of one Gaussian overlaps with another
Gaussian (i.e., such that the non-overlapping assumption would be
largely justified), nevertheless the second term in (\ref{eq:dyn.3.1-2})
can be very large despite the smallness of $R_{1}$ or $R_{2}$. It
is this latter part which is responsible for the genuinely quantum-like
nature of the average momentum, i.e., for its nonlocal nature. This
is similar in the Bohmian picture, but here given a more direct physical
meaning in that this last term refers to a difference in diffusive
currents as explicitly formulated in the last term of Eq.~(\ref{eq:3.12}).
Because of the mixing of diffusion currents from both channels, we
call this decisive term in $\mathbf{J_{\mathrm{\mathit{\mathrm{tot}}}}}$
the ``entangling current'', $\mathbf{J_{\mathit{\mathit{\mathrm{e}}}}}$
\cite{MesaPascasio.2013variable}.

\begin{figure}
\begin{minipage}[t]{0.48\columnwidth}%
\begin{flushright}
\includegraphics[scale=0.95]{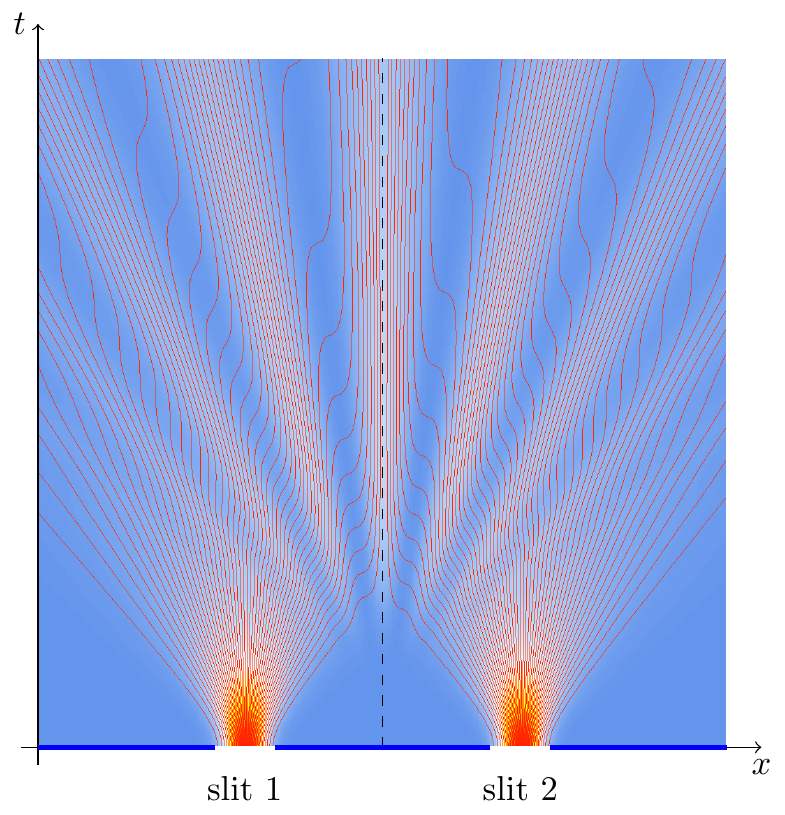}\caption{Classical computer simulation of the interference pattern with equal
slit widths: intensity distribution with increasing intensity from
white through yellow and orange, with averaged trajectories (red)
for two Gaussian slits ($v_{x,1}=v_{x,2}=0$).\label{fig:dice.1a}}

\par\end{flushright}%
\end{minipage}\hfill{}%
\begin{minipage}[t]{0.48\columnwidth}%
\begin{center}
\includegraphics[scale=0.95]{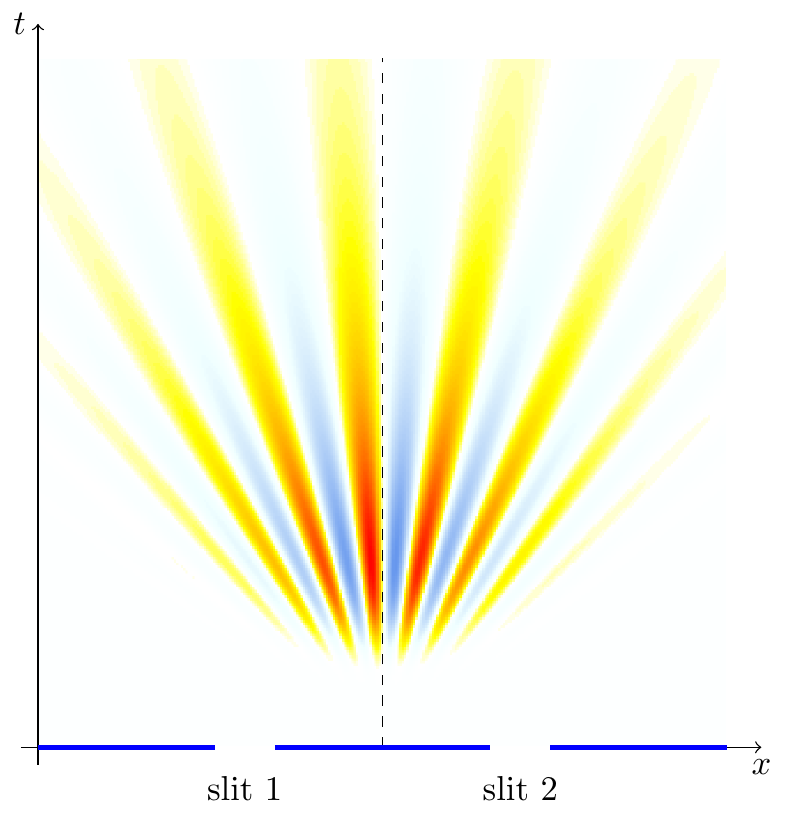}\caption{The corresponding entangling current density $\mathbf{J}_{{\rm e}}$,
i.e., the last term on the r.h.s.\ of Eq.~(\ref{eq:3.12}). As this
term is characterized by the difference of the diffusive velocities
$u_{i},$ the entangling current is responsible for the ``systemically
nonlocal'' nature of the process forming the interference pattern.\label{fig:dice.1c}}

\par\end{center}%
\end{minipage}
\end{figure}
\begin{figure}
\begin{minipage}[t]{0.48\columnwidth}%
\begin{flushright}
\includegraphics[scale=0.95]{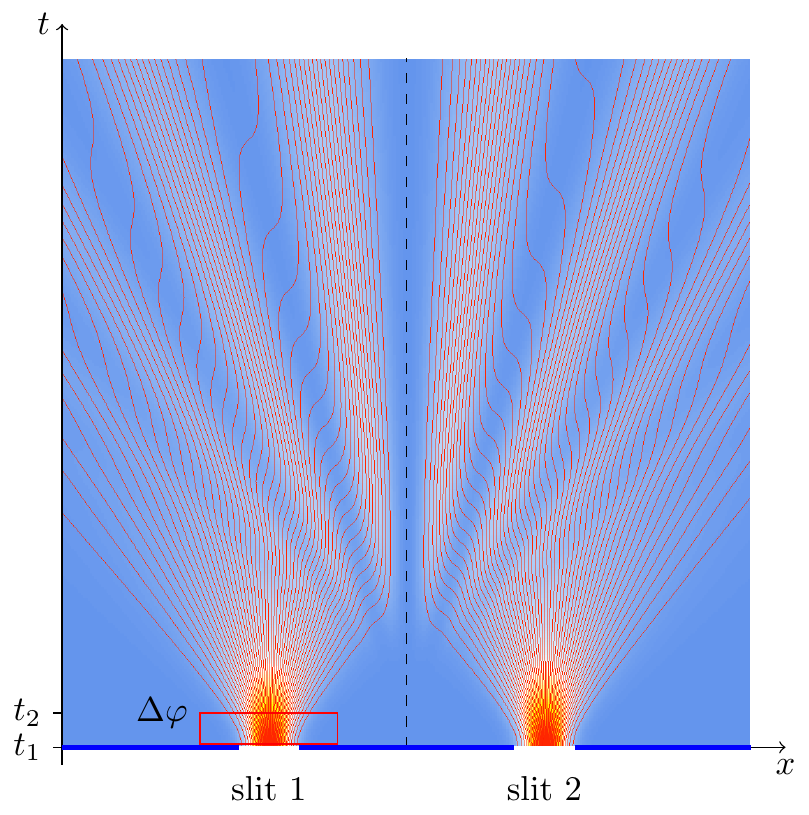}\caption{Classical computer simulation as in Fig.~\ref{fig:dice.1a}, but
with additional phase shift $\Delta\varphi=3\pi$ gradually accumulated
during the time interval between $t_{1}$ and $t_{2}$ at slit~1.\label{fig:dice.2a}}

\par\end{flushright}%
\end{minipage}\hfill{}%
\begin{minipage}[t]{0.48\columnwidth}%
\begin{center}
\includegraphics[scale=0.95]{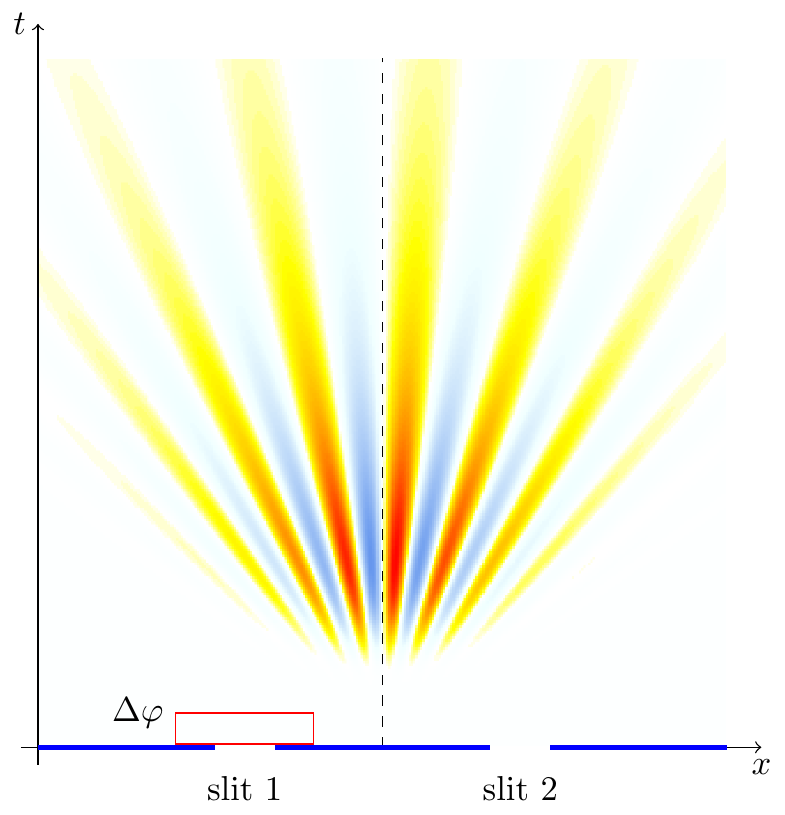}\caption{The corresponding entangling current density $\mathbf{J}_{{\rm e}}$.
Note the shift of maxima and minima in the emerging pattern, as compared
to Fig.~\ref{fig:dice.1c}.\label{fig:dice.2c}}

\par\end{center}%
\end{minipage}
\end{figure}
\begin{figure}
\begin{minipage}[t]{0.48\columnwidth}%
\begin{flushright}
\includegraphics[scale=0.95]{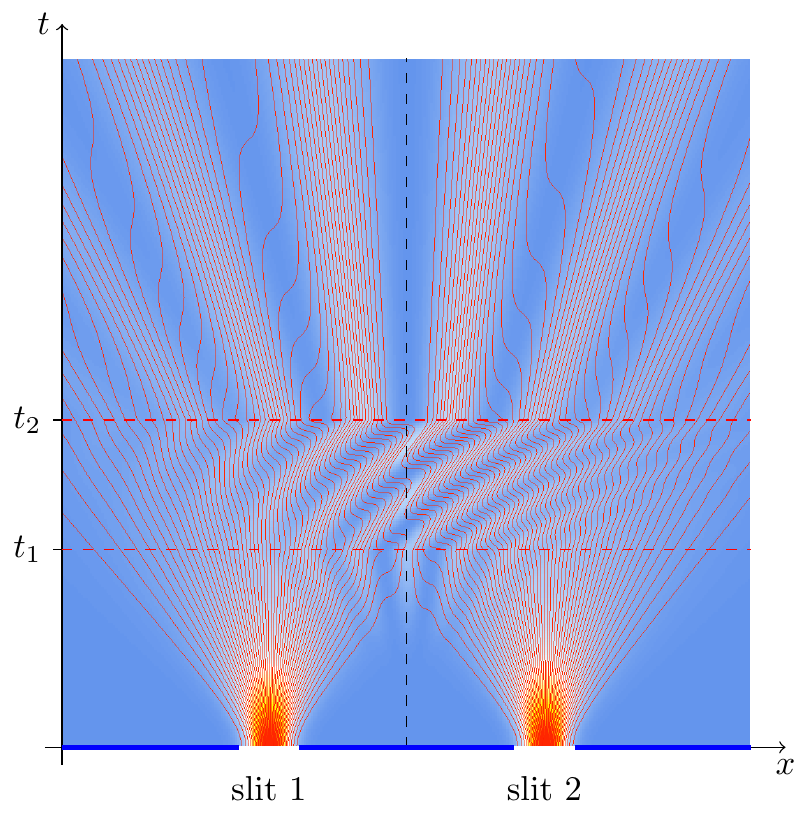}\caption{Classical computer simulation as in Fig.~\ref{fig:dice.2a}, but
with different times $t_{i}$ and with gradually accumulated additional
phase $\Delta\varphi=5\pi$. This results in the same distributions
of $P$ and $\mathbf{J}_{{\rm e}}$ for times $t>t_{2}$ and shows
the effect of the shifting of the interference fringes more clearly
than Fig.~\ref{fig:dice.2a}. Note the radically different behaviors
of the probability density related to wave-like interference on one
hand, and that of the average particle trajectories on the other.\label{fig:dice.3a}}

\par\end{flushright}%
\end{minipage}\hfill{}%
\begin{minipage}[t]{0.48\columnwidth}%
\begin{center}
\includegraphics[scale=0.95]{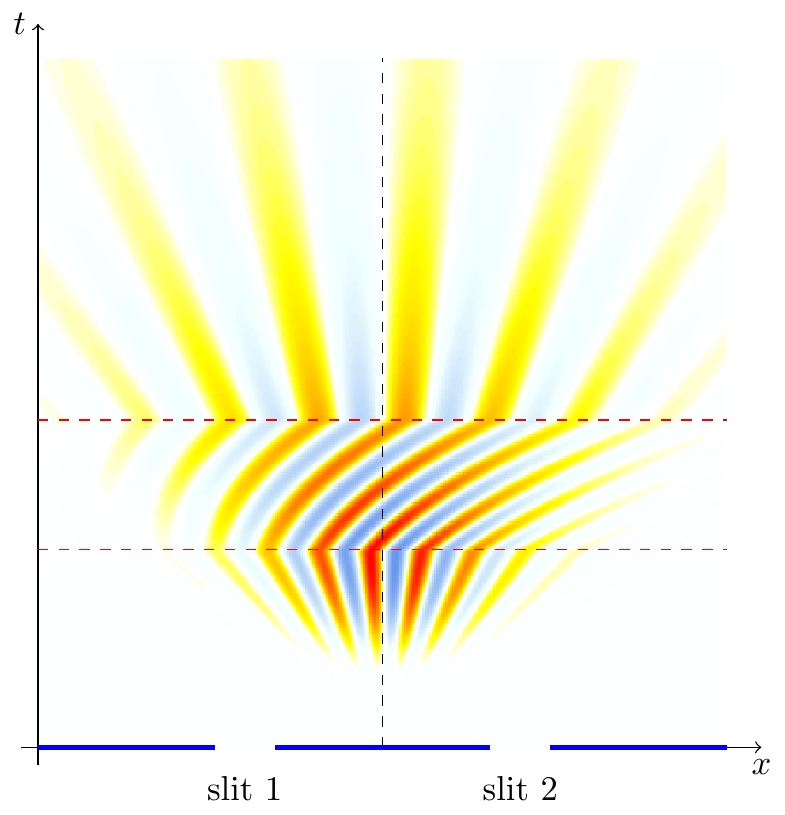}\caption{Although the currents $\mathbf{J}_{{\rm e}}$ dramatically cross the
central symmetry line separating the areas of the two slits, the average
particle trajectories (Fig.~\ref{fig:dice.3a}) strictly obey the
\emph{no crossing rule} familiar from, but not restricted to, the
deBroglie-Bohm interpretation. This is a clear demonstration of the
partial decoupling of wave and particle behavior as envisaged in our
model.\label{fig:dice.3c}}

\par\end{center}%
\end{minipage}
\end{figure}

For illustration, Figs.~\ref{fig:dice.1a}--\ref{fig:dice.3c} show
our classical computer simulations of interference and the role of
the entangling current $\mathbf{J_{\mathit{\mathit{\mathrm{e}}}}}$
in different situations. Figs.~\ref{fig:dice.1a} and \ref{fig:dice.1c}
show the emerging interference pattern and the average trajectories
without, and Figs.~\ref{fig:dice.2a} and \ref{fig:dice.2c} with,
an applied extra phase shift at one slit. To bring out the shifting
of the interference pattern more clearly, in Figs.~\ref{fig:dice.3a}
and \ref{fig:dice.3c} we apply -- mainly for didactic reasons --
the phase shift at much later times than in Figs.~\ref{fig:dice.2a}
and \ref{fig:dice.2c}. Thereby, also a decoupling of wave and particle
behaviors becomes visible.

Finally, there are substantial arguments against the non-overlapping
scenario in Tollaksen~\emph{et\,al}. Firstly, experiments by Rauch~\emph{et\,al.\ }have
shown that in interferometry interference does not only happen when
the main bulks of the Gaussians overlap, but also when a Gaussian
interferes with the off-bulk plane-wave components of the other wave
function as well \cite{Rauch.1993phase}. On a theoretical side, we
have repeatedly stressed that the diffusion processes employed in
our model can ideally be described by nonlocal diffusion wave fields
\cite{Mandelis.2001structure,Mandelis.2001diffusion-wave} which thus
require small but non-zero amplitudes across the whole experimental
setup. Our scenario may also relate to the recently developed models
of superstatistics, where a macroscopically observed emergent behavior
is the result of a superposition of two or more statistical systems
operating at vastly different space-time scales. (\cite{Jizba.2012emergence},
and Jizba and Scardigli, this volume) Moreover, and more specifically,
we have shown \cite{Groessing.2010emergence,Groessing.2011dice} that
quantum propagation can be identified with sub-quantum anomalous (i.e.,
``ballistic'') diffusion which is characterized by infinite mean
displacements $\overline{x}=\infty$ despite the finite drifts $\overline{x^{2}}=u^{2}t^{2}$
and $u<c.$ In sum, these arguments speak in favor of using small,
but non-zero amplitudes from the Gaussian of the other slit which
can interfere with the Gaussian at the particle's location.

\section{Emergence of ``Systemic Nonlocality'' from Sub-Quantum Kinematics\label{sec:Emergence-of-Dynamical}}

Let us now see how we can understand ``systemic nonlocality'' within
our sub-quantum approach. For example, we can ask the following question
in our context: what if we start with one slit only, and when the
particle might pass it we open the second slit? Let us assume for
the time being, and without restriction of generality, that the $x$--component
of the velocity $v_{x}$ is zero. Then, according to (\ref{eq:dyn.2.19}),
upon opening the second slit (with the same $\sigma$), we obtain
a term proportional to the distance $2X$ between the two slits. Thus,
there will be a shift in momentum on the particle passing the first
slit given by
\begin{equation}
\frac{\Delta p}{\hbar}=\pm\frac{1}{2}\nabla\varphi_{12}=\pm\frac{1}{2\hbar}\nabla\left(S_{1}-S_{2}\right)=\frac{\Delta p_{\mathrm{mod}}}{\hbar},\label{eq:dyn.4.1}
\end{equation}
where one effectively uses the ``modular momentum'' $p_{\mathrm{mod}}=p\textrm{ mod}\frac{h}{2X}=p-2n\pi\frac{\hbar}{2X}$
because an added or subtracted phase difference $\varphi_{12}=2n\pi$
does not change anything. In other words, by splitting $X$ into a
component $X_{n}$ providing $\varphi_{12}=2n\pi$ on the one hand,
and the modular remainder $\Delta X$ on the other hand providing
$\varphi_{12}=p_{\mathrm{mod}}x/\hbar$, one rewrites (\ref{eq:dyn.2.19})
with $X:=X_{n}+\Delta X$ (and with $v_{x}=0$ for simplicity) as
\begin{equation}
\varphi_{12}=-(X_{n}+\Delta X)x\frac{1}{D}\frac{u^{2}t}{\sigma_{0}^{2}}=:-2n\pi-\Delta Xx\frac{\hbar}{2m}\frac{t}{\sigma^{2}\sigma_{0}^{2}}\:.\label{eq:dyn.4.2}
\end{equation}

Therefore, one can further on substitute $\Delta p$ by $\Delta p_{\mathrm{mod}}$,
and one obtains a momentum shift (with the sign depending on whether
the right or left of the two slits is opened as the second one)
\begin{equation}
\Delta p_{\mathrm{mod}}=\pm\frac{\hbar}{2}\nabla\varphi_{12}=\pm\frac{\Delta X}{2}\frac{\hbar^{2}}{2m}\frac{t}{\sigma^{2}\sigma_{0}^{2}}=\pm m\Delta X\frac{D^{2}t}{\sigma^{2}\sigma_{0}^{2}}=\pm m\Delta X\frac{\dot{\sigma}}{\sigma}\:.\label{eq:dyn.4.3}
\end{equation}
For large times, one has $\dot{\sigma}\simeq\sigma/t$ such that
\begin{equation}
\Delta p_{{\rm mod}}\simeq\pm m\frac{\Delta X}{t}\:.
\end{equation}
In a separate paper \cite{MesaPascasio.2013variable} we show that,
with $\xi\left(t\right)=x-vt$ describing the location of a particle
in a Gaussian field, the action is given by
\begin{equation}
S=\int mv_{{\rm tot}}(t)\d x-\int E\d t=S=mvx+\frac{mu_{0}^{2}}{2}\left[\frac{\xi(t)}{\sigma(t)}\right]^{2}t-Et=mvx+\frac{mu_{0}^{2}}{2}\left[\frac{\xi(0)}{\sigma_{0}}\right]^{2}t-Et,
\end{equation}
with $E$ being the system's total energy. Thus, the expression $\nabla\varphi_{12}$
of (\ref{eq:dyn.4.3}) essentially refers to the gradient of the thermal
fluctuation field, with the kinetic temperature $kT=mu_{0}^{2}\left[\frac{\xi(t)}{\sigma(t)}\right]^{2}$
of what we have termed the path excitation field. 

It follows that, due to the vacuum pressure stemming from the opened
second slit, there is an \emph{emergent} ``nonlocal force'' which
does not derive from a potential but from the impinging of the second
slit's sub-quantum diffusive momenta on the particle at the first
slit. Although there is a corresponding shift in the latter's velocity
distribution, i.e.,
\begin{equation}
\frac{\partial}{\partial t}\Delta p_{\mathrm{mod}}=\pm m\Delta X\frac{1}{\sigma^{2}}\left(\sigma\ddot{\sigma}-\dot{\sigma}^{2}\right)=\pm m\Delta X\left(\frac{\ddot{\sigma}}{\sigma}-\left[\frac{\dot{\sigma}}{\sigma}\right]^{2}\right),\label{eq:dyn4.4}
\end{equation}
which reduces for large times to a gradually diminishing deceleration,
or acceleration, respectively,
\begin{equation}
\frac{\partial}{\partial t}\Delta p_{\mathrm{mod}}\simeq\mp m\Delta X\left[\frac{\dot{\sigma}}{\sigma}\right]^{2}\simeq\mp m\frac{\Delta X}{t^{2}},
\end{equation}
in the vicinity of the slits which interests us here primarily, the
expression (\ref{eq:dyn4.4}) practically vanishes identically, i.e.,
there is no additional force on the particle near the slit.

Now that we have found a possible ``nonlocal momentum transfer'',
the question immediately arises whether this would imply superluminal
signaling. To answer this question, we recall that contained in the
phase difference (\ref{eq:dyn.4.2}) is the osmotic velocity $u$,
or the diffusion-related momentum fluctuation $\delta p=mu$, respectively,
for which it holds that it must be unbiased \cite{Groessing.2004hamilton},
i.e.,
\begin{equation}
\int P\delta p\d x=0.
\end{equation}
In other words, both with regard to the directions and with regard
to the sizes, the average over all momentum fluctuations and positions
of them must vanish identically, i.e., there is a ``complete uncertainty''
of $\delta p$, just as Tollaksen~\emph{et\,al}.\ refer to a ``complete
uncertainty'' of the relative phase \cite{Tollaksen.2010quantum}.
This means that, since an experimenter can neither know where exactly
within the Gaussian distribution the particle is located nor what
quantity $\delta p=mu$ would actually be transferred in any single
run of the experiment upon opening the second slit, one can only obtain
information about the average momentum transfer $\overline{\Delta p_{\mathrm{mod}}}$
in a sufficiently large statistical sample. As the momentum-shifted
particle does not allow to extract a ``signal'' from its experiencing
that shift, i.e., as there is no way to distinguish between an unambiguous
state before and an unambiguous state after the shift, one can have
a ``nonlocal momentum transfer'' which could at best imply a ``hidden''
superluminal signaling.

Although we are therefore dealing with an epistemic indeterminism
in this case, we nevertheless have to face the problem that on an
ontic level our approach is a deterministic one (i.e., based on an
assumed stochastic, and thus essentially deterministic, sub-quantum
thermodynamics). As nonlocality can ``peacefully co-exist'' with
relativity only if one also has an underlying indeterminism, we shall
in the remainder of this paper discuss the possible consequences for
and of our model.

Firstly, the foregoing analysis was restricted to the one-dimensional
momentum transfer, i.e., in the $x$--direction normal to the forward
particle propagation. However, with the spatial components of Gaussians
being independent of each other, the corresponding dispersion and
momentum transfer processes must also hold in the other directions,
e.g., in the $y$--direction of particle propagation when considering
the two-dimensional case.

Generally, the question arises of how the calculated momentum transfer
can be nonlocal, while discarding the distances $X_{n}$ which just
provide the ineffective phase differences $\varphi_{12}=2n\pi$. Here,
the analogy to the Couder experiments provides another clue. For there,
it is clear that the modes of the standing waves emerging between
the walls of the bath container depend on the distance between the
walls. In other words, changing that distance means selecting another
set of modes. Now, it is possible to transfer this reference to the
boundary conditions also to the quantum systems we experiment with
in the laboratory. For, the whereabouts of the quanta are always restricted
to the space between sources and detectors. Changing the experimental
setup, e.g., by introducing a phase shifter, by closing a slit, or
the like, therefore always amounts to changing boundary conditions
in such a way that old modes of standing waves are substituted by
new ones to whose phases the particle/bouncer now locks in. 

In principle, this implies essentially only two options with respect
to the question of superluminal signaling (with the addition of a
third option briefly mentioned at the end of this paper): i) either,
one has a strict nonlocality, i.e., with instantaneous changes of
boundary conditions across spacelike distances, or ii) the changes
take a small, but non-zero time to become effective, thus also implying
possible superluminal effects. Option i) is an idealized one that
is argued for by using the physics of nonrelativistic oscillator-driven
diffusion wave fields as introduced above. For example, in \cite{Groessing.2010entropy}
it was shown that for the one-dimensional problem of a pulsating particle
in a box, the heat distribution turns out classically to be of the
same sinusoidal type as the quantum mechanical probability distribution.
Then, a shifting of a wall results in an instantaneous shifting of
all the nodes of the distribution just like in the quantum case. Therefore,
one has in this idealized classical scenario a nonlocality identical
with the quantum one, but originating from the interplay of oscillator-driven
diffusion wave fields with macroscopic boundaries of the experimental
setup. As mentioned, ontologically this would lead to superluminal
signaling, even if it was only a ``hidden'' one for the observers
because of the epistemic indeterminism due to the unknowability of
the exact (and, on the sub-quantum level: existing!) initial conditions. 

This leaves us with option ii) as the possibly more realistic scenario
of an emergent ``nonlocality'' that we describe as ``systemic''.
In other words, with the bouncer always oscillating harmonically with
the standing waves of the zero-point field, and with the latter being
co-determined by the experimental setup, a sudden change in the corresponding
standing wave pattern will ``practically instantaneously'' (i.e.,
during times of the order $t\sim\frac{1}{\omega}$) affect the particle.
In the case of the particle at one slit being affected by the opening
of the other slit, therefore, it does not matter how far apart the
two slits are: As long as the emerging standing wave configurations
result in a ``systemic nonlocality'' (which is a basic assumption
of our model with regard to the physical meaning of the zero-point
field), the modular remainder $\Delta X$ will make itself visible
in a shift of the particle's velocity distribution. Although framed
in a ``systemically nonlocal'' setup, therefore, the latter will
just reflect the ``local'' shift $\Delta X$ in the vicinity of
the particle. As this is accompanied by a complete uncertainty of
the relative phase, no signal can in practice be extracted from this
process. This constitutes again an explanation of the \emph{momentum
transfer} \emph{which excludes superluminal signaling in practice,
but not in principle}. However, in this scenario the simultaneity
(i.e., instantaneousness) of nonlocal effects as described by quantum
theory is lost. Still, this is very likely a problem that can be pinned
down to the problem of too idealistic applications of concepts of
relativity. In other words, the emergent quantum mechanics we envisage
may very well have to be accompanied by an \emph{emergent relativity}
as it is discussed, for example, within the paradigm of superstatistics
mentioned above. Particularly with regard to a ``double special relativity''
\cite[and Jizba and Scardigli, this volume]{Jizba.2012emergence},
classical relativity may turn out as a limiting case of a more general
approach, just as ordinary quantum theory may be a limiting case of
a ``deeper level theory'' as envisaged by \emph{emergent quantum
mechanics}.

To conclude, we can now compare the metaphysical assumptions underlying
standard quantum mechanics with those of our emergent quantum mechanics
\cite{Walleczek.2013metaphysical,Walleczek.2013detailed}. Roughly,
quantum theory may be classified via two types of approaches: i) indeterministic
and ii) deterministic ones. The first type of approaches is consistent
with standard relativity, while the second type is not. More specifically,
the first type, which one can call \emph{standard quantum mechanics}
(i.e., including interpretations of its formalism like the Copenhagen
or the deBroglie-Bohm versions), is characterized by \emph{nonlocal
indeterminism}, i.e., by a principal indeterminism that guarantees
that nonlocality cannot be used for superluminal signaling. The second,
deterministic type is characteristic for various forms of \emph{emergent
quantum mechanics}. These may either be characterized by \emph{local
determinism} with only apparent nonlocality (see, e.g., the approach
of ``stochastic electrodynamics'' as in \cite{Pena.2011,Cetto.2012quantization})
or by a \emph{determinism with ``systemic nonlocality}'', as it
characterizes one variant of our model presented here, for example. 

Within our model, we discussed in the present paper two options for
possible further elaboration. One option involves the idealized version
of nonlocal diffusion wave fields. The latter describe a classical
analogue representing nonlocality, which combined with an underlying
ontic determinism implies ``hidden'' superluminal signaling (i.e.,
despite its epistemic impossibility due to the epistemic indeterminism).
The other option is systemically nonlocal-like with only ``practical
instantaneousness'' and thus superluminal signaling, which in combination
with emergent relativity still may be constructed as free from causal
paradoxes. Finally, as mentioned, a third option can be thought of.
It would constitute a hybrid of models via the mixing of deterministic
and indeterministic elements, which is suggested as a possibility
by the assumed partial decoupling of wave and particle physics in
our model. Thus, it is at least conceivable that the deterministic
``systemic nonlocality'' described in one of our discussed options
refers only to the proposed wave-like physics, including that involving
the boundary conditions, whereas the particle-like bouncer/walker
has indeterministic degrees of freedom of its own. In this case, one
would have a ``nonlocal momentum transfer'' \emph{which excludes
superluminal signaling both in practice and in principle.} Hence,
the whole theory would effectively be indeterministic and comply with
standard relativity.

\providecommand{\href}[2]{#2}
\begingroup\raggedright
\endgroup


\begin{thebibliography}{10}

\bibitem{Groessing.2012doubleslit}
G.~Gr\"{o}ssing, S.~Fussy, J.~Mesa~Pascasio, and H.~Schwabl, ``An explanation
  of interference effects in the double slit experiment: Classical trajectories
  plus ballistic diffusion caused by zero-point fluctuations,''
  \href{http://dx.doi.org/10.1016/j.aop.2011.11.010}{{\em Ann. Phys.}
  {\bfseries 327} (2012) 421--437},
  \href{http://arxiv.org/abs/1106.5994}{{\ttfamily {arXiv}:1106.5994
  [quant-ph]}}.

\bibitem{Tollaksen.2010quantum}
J.~Tollaksen, Y.~Aharonov, A.~Casher, T.~Kaufherr, and S.~Nussinov, ``Quantum
  interference experiments, modular variables and weak measurements,''
  \href{http://dx.doi.org/10.1088/1367-2630/12/1/013023}{{\em New J. Phys.}
  {\bfseries 12} (2010) 013023},
  \href{http://arxiv.org/abs/0910.4227v1}{{\ttfamily {arXiv}:0910.4227v1
  [quant-ph]}}.

\bibitem{Aharonov.1969modular}
Y.~Aharonov, H.~Pendleton, and A.~Petersen, ``Modular variables in quantum
  theory,'' \href{http://dx.doi.org/10.1007/BF00670008}{{\em Int. J. Theor.
  Phys.} {\bfseries 2} (1969) 213--230}.

\bibitem{Aharonov.1970deterministic}
Y.~Aharonov, H.~Pendleton, and A.~Petersen, ``Deterministic quantum
  interference experiments,'' \href{http://dx.doi.org/10.1007/BF00672451}{{\em
  Int. J. Theor. Phys.} {\bfseries 3} (1970) 443--448}.

\bibitem{Spence.2012experimental}
S.~Spence and A.~Parks, ``Experimental evidence for a dynamical non-locality
  induced effect in quantum interference using weak values,''
  \href{http://dx.doi.org/10.1007/s10701-011-9596-6}{{\em Found. Phys.}
  {\bfseries 42} (2012) 803--815},
  \href{http://arxiv.org/abs/1010.3289v1}{{\ttfamily {arXiv}:1010.3289v1
  [quant-ph]}}.

\bibitem{Groessing.2008vacuum}
G.~Gr\"{o}ssing, ``The vacuum fluctuation theorem: Exact schr\"{o}dinger
  equation via nonequilibrium thermodynamics,''
  \href{http://dx.doi.org/10.1016/j.physleta.2008.05.007}{{\em Phys. Lett. A}
  {\bfseries 372} (2008) 4556--4563},
  \href{http://arxiv.org/abs/0711.4954v2}{{\ttfamily {arXiv}:0711.4954v2
  [quant-ph]}}.

\bibitem{Groessing.2009origin}
G.~Gr\"{o}ssing, ``On the thermodynamic origin of the quantum potential,''
  \href{http://dx.doi.org/10.1016/j.physa.2008.11.033}{{\em Physica A}
  {\bfseries 388} (2009) 811--823},
  \href{http://arxiv.org/abs/0808.3539v1}{{\ttfamily {arXiv}:0808.3539v1
  [quant-ph]}}.

\bibitem{Groessing.2010emergence}
G.~Gr\"{o}ssing, S.~Fussy, J.~Mesa~Pascasio, and H.~Schwabl, ``Emergence and
  collapse of quantum mechanical superposition: Orthogonality of reversible
  dynamics and irreversible diffusion,''
  \href{http://dx.doi.org/10.1016/j.physa.2010.07.017}{{\em Physica A}
  {\bfseries 389} (2010) 4473--4484},
  \href{http://arxiv.org/abs/1004.4596}{{\ttfamily {arXiv}:1004.4596
  [quant-ph]}}.

\bibitem{Groessing.2011dice}
G.~Gr\"{o}ssing, S.~Fussy, J.~Mesa~Pascasio, and H.~Schwabl, ``Elements of
  sub-quantum thermodynamics: quantum motion as ballistic diffusion,''
  \href{http://dx.doi.org/10.1088/1742-6596/306/1/012046}{{\em J. Phys.: Conf.
  Ser.} {\bfseries 306} (2011) 012046},
  \href{http://arxiv.org/abs/1005.1058}{{\ttfamily {arXiv}:1005.1058
  [physics.gen-ph]}}.

\bibitem{Groessing.2011explan}
G.~Gr\"{o}ssing, J.~Mesa~Pascasio, and H.~Schwabl, ``A classical explanation of
  quantization,'' \href{http://dx.doi.org/10.1007/s10701-011-9556-1}{{\em
  Found. Phys.} {\bfseries 41} (2011) 1437--1453},
  \href{http://arxiv.org/abs/0812.3561}{{\ttfamily {arXiv}:0812.3561
  [quant-ph]}}.

\bibitem{Grossing.2012quantum}
G.~Gr\"{o}ssing, S.~Fussy, J.~Mesa~Pascasio, and H.~Schwabl, ``The quantum as
  an emergent system,''
  \href{http://dx.doi.org/10.1088/1742-6596/361/1/012008}{{\em J. Phys.: Conf.
  Ser.} {\bfseries 361} (2012) 012008},
  \href{http://arxiv.org/abs/1205.3393}{{\ttfamily {arXiv}:1205.3393
  [quant-ph]}}.

\bibitem{Couder.2005}
Y.~Couder, S.~Proti\`{e}re, E.~Fort, and A.~Boudaoud, ``Dynamical phenomena:
  Walking and orbiting droplets,''
  \href{http://dx.doi.org/10.1038/437208a}{{\em Nature} {\bfseries 437} (2005)
  208--208}.

\bibitem{Couder.2006single-particle}
Y.~Couder and E.~Fort, ``Single-particle diffraction and interference at a
  macroscopic scale,''
  \href{http://dx.doi.org/10.1103/PhysRevLett.97.154101}{{\em Phys. Rev. Lett.}
  {\bfseries 97} (2006) 154101}.

\bibitem{Couder.2010walking}
Y.~Couder, A.~Boudaoud, S.~Proti\`{e}re, and E.~Fort, ``Walking droplets, a
  form of wave-particle duality at macroscopic scale?,''
  \href{http://dx.doi.org/10.1051/epn/2010101}{{\em Europhys. News} {\bfseries
  41} (2010) 5}.

\bibitem{Couder.2012probabilities}
Y.~Couder and E.~Fort, ``Probabilities and trajectories in a classical
  wave-particle duality,''
  \href{http://dx.doi.org/10.1088/1742-6596/361/1/012001}{{\em J. Phys.: Conf.
  Ser.} {\bfseries 361} (2012) 012001}.

\bibitem{Bohm.1993undivided}
D.~Bohm and B.~J. Hiley, {\em The undivided universe: An ontological
  interpretation of quantum theory}.
\newblock Routledge, London, 1993.

\bibitem{Holland.1993}
P.~R. Holland, {\em The Quantum Theory of Motion: An account of the de
  Broglie-Bohm causal interpretation of quantum mechanics}.
\newblock Cambridge University Press, Cambridge, 1993.

\bibitem{DeBroglie.1960book}
L.~V. P.~R. de~Broglie, {\em Non-Linear Wave Mechanics: A Causal
  Interpretation.}
\newblock Elsevier, Amsterdam, 1960.

\bibitem{Hall.2002schrodinger}
M.~J.~W. Hall and M.~Reginatto, ``Schr\"{o}dinger equation from an exact
  uncertainty principle,''
  \href{http://dx.doi.org/10.1088/0305-4470/35/14/310}{{\em J. Phys. A: Math.
  Gen.} {\bfseries 35} (2002) 3289--3303}.

\bibitem{Fort.2010path-memory}
E.~Fort, A.~Eddi, A.~Boudaoud, J.~Moukhtar, and Y.~Couder, ``Path-memory
  induced quantization of classical orbits,''
  \href{http://dx.doi.org/10.1073/pnas.1007386107}{{\em {PNAS}} {\bfseries 107}
  (2010) 17515--17520}.

\bibitem{Mandelis.2001structure}
A.~Mandelis, L.~Nicolaides, and Y.~Chen, ``Structure and the
  {Reflectionless/Refractionless} nature of parabolic diffusion-wave fields,''
  \href{http://dx.doi.org/10.1103/PhysRevLett.87.020801}{{\em Phys. Rev. Lett.}
  {\bfseries 87} (2001) 020801}.

\bibitem{Sanz.2008trajectory}
A.~S. Sanz and S.~Miret-Art\'{e}s, ``A trajectory-based understanding of
  quantum interference,''
  \href{http://dx.doi.org/10.1088/1751-8113/41/43/435303}{{\em J. Phys. A:
  Math. Gen.} {\bfseries 41} (2008) 435303},
  \href{http://arxiv.org/abs/0806.2105}{{\ttfamily {arXiv}:0806.2105
  [quant-ph]}}.

\bibitem{MesaPascasio.2013variable}
J.~Mesa~Pascasio, S.~Fussy, H.~Schwabl, and G.~Gr\"{o}ssing, ``Modeling quantum
  mechanical double slit interference via anomalous diffusion: independently
  variable slit widths,''
  \href{http://dx.doi.org/10.1016/j.physa.2013.02.006}{{\em Physica A, In
  Press} (2013) }.

\bibitem{Rauch.1993phase}
H.~Rauch, ``Phase space coupling in interference and {EPR} experiments,''
  \href{http://dx.doi.org/10.1016/0375-9601(93)90270-A}{{\em Phys. Lett. A}
  {\bfseries 173} (1993) 240--242}.

\bibitem{Mandelis.2001diffusion-wave}
A.~Mandelis, {\em Diffusion-wave fields: Mathematical methods and Green
  functions}.
\newblock Springer, New York, {NY}, 2001.

\bibitem{Jizba.2012emergence}
P.~Jizba and F.~Scardigli, ``Emergence of special and doubly special
  relativity,'' \href{http://dx.doi.org/10.1103/PhysRevD.86.025029}{{\em Phys.
  Rev. D} {\bfseries 86} (2012) 025029},
  \href{http://arxiv.org/abs/1105.3930}{{\ttfamily {arXiv}:1105.3930
  [hep-th]}}.

\bibitem{Groessing.2004hamilton}
G.~Gr\"{o}ssing, ``From classical hamiltonian flow to quantum theory:
  Derivation of the schr\"{o}dinger equation,''
  \href{http://dx.doi.org/10.1023/B:FOPL.0000035669.03595.ce}{{\em Found. Phys.
  Lett.} {\bfseries 17} (2004) 343--362},
  \href{http://arxiv.org/abs/quant-ph/0311109v2}{{\ttfamily
  quant-ph/0311109v2}}.

\bibitem{Groessing.2010entropy}
G.~Gr\"{o}ssing, ``Sub-quantum thermodynamics as a basis of emergent quantum
  mechanics,'' \href{http://dx.doi.org/10.3390/e12091975}{{\em Entropy}
  {\bfseries 12} (2010) 1975--2044}.

\bibitem{Walleczek.2013metaphysical}
J.~Walleczek, {\em to be published in Found. Phys.} (2013) .

\bibitem{Walleczek.2013detailed}
J.~Walleczek and G.~Gr\"{o}ssing, {\em to be published in Found.
  Phys.} (2013) .

\bibitem{Pena.2011}
L.~de~la Pe\~{n}a, A.~Vald\'{e}s-Hern\'{a}ndez, A.~Cetto, and H.~Fran\c{c}a,
  ``Genesis of quantum nonlocality,''
  \href{http://dx.doi.org/10.1016/j.physleta.2011.02.024}{{\em Phys. Lett. A}
  {\bfseries 375} (2011) 1720--1723}.

\bibitem{Cetto.2012quantization}
A.~M. Cetto and L.~de~la Pe\~{n}a, ``Quantization as an emergent phenomenon due
  to matter-zeropoint field interaction,''
  \href{http://dx.doi.org/10.1088/1742-6596/361/1/012013}{{\em J. Phys.: Conf.
  Ser.} {\bfseries 361} (2012) 012013}.

\end{thebibliography}
\end{document}